\documentclass[12pt,preprint,longabstract]{aastex}
\newcommand{\text}[1]{\ensuremath{\mathrm{#1}}}
\usepackage{units}
\usepackage{epsfig}

\begin{document}
\title{Search for Neutrinos from GRB 080319B at Super-Kamiokande}

\author{
E. Thrane\altaffilmark{20},
K. Abe\altaffilmark{1},
Y. Hayato\altaffilmark{1,33},
T. Iida\altaffilmark{1},
M. Ikeda\altaffilmark{1},
J. Kameda\altaffilmark{1},
K. Kobayashi\altaffilmark{1},
Y. Koshio\altaffilmark{1,33},
M. Miura\altaffilmark{1},
S. Moriyama\altaffilmark{1,33}
% ...
}

\affil{The Super-Kamiokande Collaboration}

\altaffiltext {1}{Kamioka Observatory, Institute for Cosmic Ray Research, The University of Tokyo, Hida, Gifu 506-1205, Japan}
% ...
\altaffiltext{20}{Department of Physics and Astronomy, University of Minnesota, Minneapolis, MN, 55455, USA}
% ...
\altaffiltext{33}{Institute for the Physics and Mathematics of the Universe (IPMU), The University of Tokyo, Kashiwa, Chiba 277-8568, Japan}
% ...

\begin{abstract}
We perform a search for neutrinos coincident with GRB 080319B---the brightest GRB observed to date---in a $\pm\unit[1,000]{s}$ window.
No statistically significant coincidences were observed and we thereby obtain an upper limit on the fluence of neutrino-induced muons from this source.
From this we apply reasonable assumptions to derive a limit on neutrino fluence from the GRB. 
\end{abstract}

%\pacs{95.55.Vj, 95.85.Ry}
\maketitle

\section{Introduction}   
On 2008 March 19 at 06:12:49~UT (15:12:49 Japan Standard Time) the Swift Burst Alert Telescope recorded a gamma-ray burst (designated GRB~080319B) at equatorial coordinates $(\text{ra}=217^\circ56^{\prime\prime},\text{dec}=+36^\circ18^{\prime\prime})$ \citep{GRBNetwork}.
This GRB was simultaneously detected with the Konus $\gamma$-ray detector on board the Wind satellite \citep{konus}, and subsequently by detectors operating at lower bandwidths including ``Pi of the Sky'' \citep{pi}, TORTORA \citep{tortora}, the Liverpool Telescope \citep{liverpool}, the Faulkes Telescope North, Gemini-North, the Hobby-Eberly Telescope, and the Very Large Telescope.
The burst had a duration of $t_{90}>\unit[50]{s}$ \citep{racusin} (making it a ``long'' GRB,) with a fluence of $\unit[6.23\pm0.13\times10^{-4}]{ergs\,cm^{-2}}$ in the $\unit[20]{keV}$-$\unit[7]{MeV}$ band \citep{nature}.
It is believed that such long GRBs may be due to hypernovae from collapsing Wolf-Rayet stars \citep{woosley}.
This corresponds to an isotropic-equivalent energy release of $\approx\unit[10^{54}]{ergs}$ at a luminosity distance of $\unit[1.9\times10^{28}]{cm}$ \citep{nature}.
It occurred at a redshift of $z=0.937$ \citep{nature}.
This GRB is remarkable as it is the most distant astronomical object ever visible to the naked eye \citep{Naeye}.
Some GRB models predict large fluxes of neutrinos over a wide range of energies \citep{LearnedMannheim,Meszaros,Eichler}, and so in this article we describe a search for neutrinos at Super-Kamiokande coincident with photonic observations of GRB~080319B.

\section{Super-Kamiokande}
Super-Kamiokande is a water Cherenkov detector located within Mt. Ikeno in central Japan, under $\unit[2,700]{meters}$ water equivalent rock overburden.
It has a cylindrical design, holds $\unit[50]{kilotons}$ of water, and is divided into two optically separated sections by a structural framework that supports photomultiplier tubes (PMTs).
Super-Kamiokande has an inner detector (ID) equipped with 11,146 $\unit[50]{cm}$ PMTs aimed inward (at the time considered here,) and an outer detector (OD) volume instrumented with 1,885 $\unit[20]{cm}$ PMTs aimed outward and equipped with wavelength-shifting plastic plates.
The OD functions primarily as a veto counter, tagging charged particles that enter or exit the ID.
Within the ID we define a central $\unit[22.5]{kiloton}$ fiducial volume within which detector response is expected to be uniform \citep{Fukuda:2002uc}.

Neutrino events (interactions recorded in the detector) with total deposited energy above $\approx\unit[100]{MeV}$ are overwhelmingly due to atmospheric neutrinos (decay products from high-energy cosmic ray interactions in the upper atmosphere.)
The Earth is essentially transparent to such neutrinos up to energies on the order of $\unit[100]{TeV}$.

There are several types of events observed at Super-Kamiokande categorized according to their energy and/or topology.
The low-energy sample is typically used to study solar neutrinos and to search for core-collapse supernova neutrinos via elastic $\nu-e$ scattering for all flavors of neutrinos and $\bar{\nu}_e+p^+\rightarrow e^+ +n$ for anti-electron-neutrinos.
The low-energy event selection algorithm---designed to reduce background from cosmic-ray muons, decay electrons, penetrating muon-induced spallation events, and radioactivity from the surrounding rock---is described in \cite{Parker,Hosaka}.
Low-energy events associated with core-collapse supernovae produced in hypernovae scenarios are not thought to be beamed like high-energy neutrinos, making low-energy neutrino detection difficult at these cosmological distances.
If we assume that $m_\nu=\unit[1]{eV}$ and $E_\nu=\unit[10]{MeV}$, then the relativistic delay for neutrinos from GRB~080319B would be on the order of $\unit[180]{s}$ \citep{Hong}, which falls within the $\pm \unit[1,000]{s}$ search window discussed below.

Fully-contained (FC) neutrino events are those where interaction products are observed in the ID with no significant correlated activity in the OD, while partially-contained (PC) events are those where some interaction products exit the ID.

Upward-going muon (upmu) events are those where a penetrating particle traveling in the upward direction enters and either stops in or passes through the detector, and are attributed to muons produced by neutrino interactions in the surrounding rock.
We require that upmus traverse at least $\unit[7]{m}$ in the ID if exiting or that they exhibit the equivalent energy loss ($\unit[1.6]{GeV}$) for that range if stopping.
Using Monte Carlo data, we measure the efficiency of the upmu reduction algorithm by calculating the ratio of events that are both true upmus and also tagged as upmus, to events that are true upmus (and may or may not be tagged as upmus.)
The upmu reduction efficiency at the time of GRB~080319B was $\approx98\%$ \citep{thrane}.
A variety of astrophysical results obtained with upmus can be found in \citet*{SKGRB,shantanu,Kristine,WIMPs}.

In general terms, FC, PC, and upmu events represent successively higher energy samples of neutrino interactions, ranging from $E_\nu\sim\unit[100]{MeV}$ and corresponding to $>\unit[200]{photoelectrons}$ (pe) for the lowest energy FC events to above $E_\nu\sim\unit[1]{TeV}$ ($<\unit[2\times10^6]{pe}$) for the highest energy upmus.
Each category of event can be considered a coarse energy bin so long as it is understood that the bins overlap significantly.
Further details regarding the Super-Kamiokande detector design, operation, calibration, and data reduction can be found in \citet*{Fukuda:2002uc,Ashie:2005ik}.

The effective area of the detector depends upon zenith angle and varies between $\unit[960-1,300]{m^2}$.
For each zenith angle, we determine effective area by performing a 2D projection of the cylindrical detector onto a plane, subject to the constraint that a particle passing through the detector must be able to traverse at least $\unit[7]{m}$ through the ID.

\section{Search Method}
Following a method developed in a previously published GRB search \citep{SKGRB}, we employ $\pm\unit[1,000]{s}$ timing window centered on the beginning of photonic observations of GRB~080319B and look for coincident neutrino events.
This window size allows for any reasonable delay, positive or negative, between neutrino emission and photonic emission, while still providing a very small likelihood of random coincidences due to atmospheric neutrinos.
We consider all four categories of atmospheric neutrino events---low-energy, FC, PC, and upmu---but the strongest limits come from the upmu dataset.

To eliminate cosmic ray background, we require upmu events to come from below the horizon, which means that the upmu sample is only sensitive to declinations less than $+54^\circ$.
Fortunately, GRB~080319B occurred $17^\circ$ below the horizon, with a local azimuthal angle of $\phi=32^\circ$.

The upmu reduction algorithm is divided into several stages designed to filter out downward-going cosmic-ray muons, which sometimes masquerade as upmus \citep{Ashie:2005ik}, as well as ultra-high-energy (UHE) events, (with $>\unit[1.75\times10^6]{pe}$ in the ID,) that are analyzed separately for astronomy studies.
UHE background events occur at a rate of about $\unit[30]{day^{-1}}$.
{\it Upward-going} UHE events, (which can be attributed to neutrinos,) are very rare and occur at a rate of $\unit[2.7\times10^{-4}]{day^{-1}}$ \citep*{molly}.

\begin{figure}
  \epsscale{1.00}
  \plotone{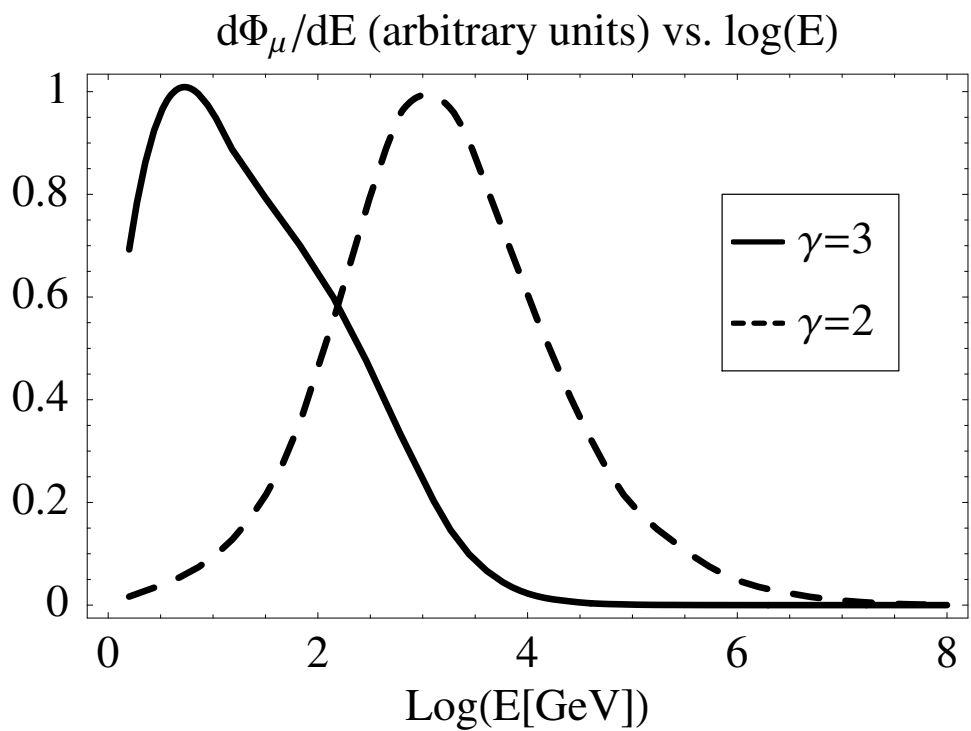}
  \caption{The upmu spectra for a source with spectral indices $\gamma=2$ (dashed) and $\gamma=3$ (solid). Analysis cuts impose an energy threshold $E_\mu^\text{min}\ge\unit[1.6]{GeV}$. \label{fig:upmuspec}}
\end{figure}

Currently favored GRB models predict a hard $d\Phi_\nu/dE\propto E^{-2}$ spectrum \citep{LearnedMannheim,Meszaros}.
Combining this with the roughly linear energy dependence of the neutrino-nucleon cross section, plus the fact that the long range of high-energy upmus makes the effective detector volume increase with energy, we expect a peaked spectrum for observed muons, as shown in Figure~\ref{fig:upmuspec}.
The position of the peak depends upon how hard the GRB neutrino spectrum turns out to be, but the most likely range for spectral indices ($\gamma = 2 \sim 3$) puts the peak in the energy region spanned by Super-Kamiokande's upmu events.

The detector's angular resolution---defined such that 68\% of events have an angular separation between true and fit muon direction that is smaller than the resolution---is about $1^\circ$ at this zenith angle.
At energies above $\unit[1]{TeV}$ the typical angular separation between a neutrino and its daughter muon is less than $1^\circ$ \citep{LearnedMannheim}.
The effective area at this zenith angle is $\approx\unit[1,270]{m^2}$.

The rates of post-reduction atmospheric and solar background events are summarized in Table~\ref{tab:bknd}.
Since the background rates are so low, a coincidence in time provides a strong signal of a correlation between the observed neutrino and the photonic detection.
(The probability of an accidental upmu coincidence from an atmospheric neutrino is about 4\%.)
It is also possible to employ a cut on the reconstructed upmu direction in relation to the GRB direction, but this was not necessary as there were no upmu coincidences in time.

\begin{deluxetable}{cccc}
  \tabletypesize{\scriptsize}
  \tablecaption{Background rates and events observed in window. \label{tab:bknd}}
  \tablewidth{0pt}
  \tablehead{
    \colhead{type} & \colhead{background ($\text{day}^{-1}$)} & \colhead{...in $\unit[\pm1,000]{s}$ window} & \colhead{observed in window}
  }
    \startdata
    low-energy & $390$ & $9$ & $11\pm3$ \\
    FC & $8$ & 0.2 & 0 \\
    PC & $0.6$ & 0.01 & 0 \\
    upmu & $1.6$ & 0.04 & 0 \\
    UHE & $2.7\times10^{-4}$ & $6\times10^{-6}$ & 0 \\
    \enddata
\end{deluxetable}

\section{Results}
No statistically significant coincidences were observed in any of the four categories (see Table~\ref{tab:bknd}).
Therefore we set an upper limit on neutrino-induced muon fluence, the effective neutrino flux integrated over the duration of the burst, which is the appropriate physical quantity to describe transient sources.
We find that the upper limit at 90\% on the fluence of neutrino-induced muons from GRB~080319B is $\Phi^{90\%}_\mu=\unit[1.96\pm0.04\times10^{-7}]{cm^{-2}}$.
(The $\approx2\%$ uncertainty is due primarily to uncertainty in our track length reconstruction algorithm and in our live time calculation.)

%We find the low-energy event rate for events with $\unit[5]{MeV}<E<\unit[80]{MeV}$ during the $\pm\unit[1,000]{s}$ window to be $\unit[0.0055\pm0.0017]{s^{-1}}$, which is consistent with the normal background rate of $\approx\unit[0.0045]{s^{-1}}$.

\section{Neutrino Fluence Limit}
Extrapolating from upmu fluence limits to neutrino fluence limits requires making an assumption about the nature of the source spectrum.
We estimate the muon neutrino flux at the detector (after neutrinos have oscillated from their source distribution.)
Since a wide variety of models predict a power law spectrum with a spectral index of about $\gamma=2$, we assume $d\Phi_\nu/dE_\nu\propto E^{-\gamma}$ and consider the cases of $\gamma=2$ and $\gamma=3$.

Upmu fluence can be related to neutrino fluence as in Equation~\ref{eq:nu2mu}.
\begin{eqnarray}\label{eq:nu2mu}
  \lefteqn{ \Phi_\mu(>E_\mu^\text{min}) = \int_{E_\mu^\text{min}}^\infty dE_\nu } \nonumber \\
  && \left[ P(E_\nu, E_\mu^\text{min}) S(z,\, E_\nu) \, \frac{d\Phi_\nu}{dE_\nu} \right]
\end{eqnarray}
Here $\Phi_\mu(>E_\mu^\text{min})$ is the fluence of upmus with energies above the minimum upmu energy of $E_\mu^\text{min}\equiv\unit[1.6]{GeV}$.
The quantity $P(E_\nu,\, E_\mu^\text{min})$ is the probability that a neutrino with energy $E_\nu$ creates a muon with energy greater than $E_\mu^\text{min}$ and $S(z, E_\nu)$ is the Earth's shadow factor.
We use cross sections from the GRV94 parton distribution function \citep{Gluck} and the muon range is determined using \cite{Lipari,Reno}.

\begin{eqnarray}\label{eq:nu2muProb}
  \lefteqn{  P(E_\nu,\, E_\mu^\text{min})= N_A \int_0^{E_\nu}dE_\mu } \nonumber \\
&& \left[ \frac{d\sigma_{CC}}{dE_\mu}(E_\mu,E_\nu) R(E_\mu,E_\mu^\text{min}) \right]
\end{eqnarray}
Here $d\sigma_{CC}/dE_\mu(E_\mu,\,E_\nu)$ is the charged current component of the neutrino-nucleon cross section; (neutral current interactions do not produce penetrating muons.)
The quantity $R(E_\mu,\,E_\mu^\text{min})$, meanwhile, is the average range in rock for a muon with an initial energy of $E_\mu$ and a final energy greater than $E_\mu^\text{min}$.
The quantity $N_A$ is the water equivalent Avogadro's number (scaled by the density of water.)
\begin{equation}\label{eq:shadow}
  S(z,\,E_\nu)=e^{-l_\text{col}(z)\,\sigma(E_\nu)\,N_A}
\end{equation}
Here $l_\text{col}(z)$ is the angle-dependent column depth along the neutrino path.
Column depth is calculated in \cite{Gandhi96} using the ``Preliminary Earth Model.''
The results of this calculation are summarized in Table~\ref{tab:nufluence}.

\begin{deluxetable}{lcr}
  \tabletypesize{\scriptsize}
  \tablecaption{Limits at 90\% CL on the fluence of muon neutrinos and antineutrinos from GRB~080319B given different spectral indices and derived with the upmu sample. \label{tab:nufluence}}
  \tablewidth{0pt}
  \tablehead{
    \colhead{spectral index} & \colhead{$\Phi_\nu$ ($\text{cm}^{-2}$)} & \colhead{$\bar{\Phi}_\nu\nu$ ($\text{cm}^{-2}$)}
    }
\startdata
$\gamma=2$ & $16\pm1.7$ & $22\pm2.4$ \\
$\gamma=3$ & $6.6\pm0.7\times10^3$ & $10\pm1.1\times10^3$ \\
\enddata
\end{deluxetable}

In order to derive fluence limits for low-energy and FC/PC events we use Equation~\ref{FCPCFluence}, which differs from Equation~\ref{eq:nu2mu} because low-energy and FC/PC interactions must occur in the water fiducial volume, while upmu interactions occur in the rock outside the detector.
\begin{equation}\label{FCPCFluence}
  \Phi_\text{FC/PC}=\frac{N_{90}}{\sum_i \, N^i_T\,\int dE_\nu \, \sigma^i(E_\nu) \, \epsilon(E_\nu) \, E_\nu^{-\gamma}}
\end{equation}
Here $N_{90}$ is the 90\% CL limit on the total number of FC/PC neutrino interactions in the window and $N^i_T$ is the number of interaction targets of type $i$ where $i$ can designate neutrons, protons, or electrons.
The quantity $\sigma^i(E_\nu)$ is the cross section for neutrinos to interact with particles of type $i$ and $\epsilon(E_\nu)$ is the detector efficiency as a function of neutrino energy.
The low-energy limits use cross sections from \cite{Strumia}.
The FC/PC limits and low-energy limits are presented in Table~\ref{tab:FCPC}.

\begin{deluxetable}{lr}
  \tabletypesize{\scriptsize}
  \tablecaption{Limits at 90\% CL on the fluence of neutrinos from GRB~080319B given a spectral index of $\gamma=2$ and derived with the FC/PC data sample (top), with the low-energy sample (middle), and with UHE data (bottom). \label{tab:FCPC}}
  \tablewidth{0pt}
  \tablehead{
    \colhead{species} & \colhead{$\Phi_\nu$ ($\text{cm}^{-2}$)}
  }
  \startdata
  & {\bf from FC/PC only} \\
  $\nu_\mu$ & $3.5\times10^4$ \\       
  $\bar{\nu}_\mu$ & $7.6\times10^4$ \\ 
  $\nu_e$ & $4.3\times10^4$ \\         
  $\bar{\nu}_e$ & $7.6\times10^4$ \\   
  & {\bf from low-energy only} \\
  $\nu_\mu$ & $5.1\times10^7$ \\
  $\nu_e$ & $8.0\times10^6$ \\
  $\bar{\nu}_e$ & $3.0\times10^4$ \\
  & {\bf from UHE only} \\
  $\nu_\mu$ & $1.2\times10^3$ \\
  $\bar{\nu}_\mu$ & $1.6\times10^3$ \\
  \enddata
\end{deluxetable}

\section{Systematic Uncertainty}
Uncertainty in neutrino-nucleon cross section creates systematic uncertainty in the calculated neutrino fluence.
In Equation~\ref{eq:nu2mu}, we utilize a table of neutrino-nucleon cross section, which has an associated uncertainty of $\approx10\%$ at the upmu energy scale \citep{Gandhi96}.
To assess how this affects the neutrino fluence uncertainty, we carry out the calculation using $\Phi_+\equiv\Phi(\sigma\rightarrow110\%\,\sigma)$ as well as $\Phi_-\equiv\Phi(\sigma\rightarrow90\%\,\sigma)$ and estimate the uncertainty as half the difference.
For upmus, an additional uncertainty arises from cross section uncertainty in the Earth shadow.
We add these two uncertainties in quadrature and find the total uncertainty in $\Phi_\nu$ to be 11\%.

\section{Comparison with Other Results}
In \cite{SKGRB} a set of 1,454 GRBs from the BATSE catalog was used to test for coincident neutrinos at Super-Kamiokande.
Assuming a spectral index of $\gamma=2$, a 90\% limit was obtained for the average GRB fluence of $\Phi^{90\%}_\nu<\unit[0.038]{cm^{-2}}$ for $\nu_\mu$ ($\unit[0.050]{cm^{-2}}$ for $\bar{\nu}_\mu$.)
Since this limit is derived from the summed fluence of 1,454 GRBs it is, of course, significantly lower than the one quoted here and obtained from the consideration of a single GRB.
(The limit quoted here, however, is better than the ``average'' single-burst limit from \cite{SKGRB}, $\Phi_\nu<\unit[55]{\text{cm^{-2}}}$.)

All the same, the importance of the present result is magnified by the fact that GRB~080319B was so exceptionally bright.
Further, studies of burst-to-burst fluctuations suggest that variations in the Lorentz factor characterizing GRB shockwaves can change the neutrino fluence by roughly four orders of magnitude \citep{HalzenHooper}.
Likewise, modeling of individual GRBs reveals a similarly wide range of predicted fluences \citep{Guetta}.

The best limits on average GRB fluence currently come from the AMANDA experiment, which reports an upper limit of $\unit[1.4\times10^{-5}]{cm^{-2}}$ for neutrino energies between $\unit[250-10^7]{GeV}$ and assuming a spectral index of $\gamma=2$ \citep{AMANDA}.
AMANDA has also searched for neutrinos from a single GRB (GRB~030329) and set differential fluence limits (for several models) in the range of $d\Phi/d\log{E}=\unit[1-6]{GeVcm^{-2}}$ \citep*{Stamatikos}.

\section{Conclusions}
We have looked for neutrinos coincident in time with GRB~080319B---the brightest GRB observed to date.
We found no significant coincidences, and so we set an upper limit on the fluence of neutrinos from this object.

\section{Acknowledgments}
Data on muon range and neutrino cross sections were graciously provided by M. Reno.
The authors gratefully acknowledge the cooperation of the Kamioka Mining and Smelting Company.
Super-Kamiokande has been built and operated from funds provided by the Japanese Ministry of Education, Culture, Sports, Science and Technology as well as the U.S. Department of Energy and the U.S. National Science Foundation.
Some participants have been supported by funds from the Korean Research Foundation (BK21, BRP) and the Korea Science and Engineering Foundation.

\bibliographystyle{apj}
\bibliography{ms}

\end{document}